# Search for Rare and Forbidden Decays of D mesons to Final States with Electrons

**David H Miller**[*][†]
*Purdue University, West Lafayette, USA*
*E-mail:* miller@physics.purdue.edu

Using the CLEO-c detector and a data sample of 0.8 million $D^+D^-$ pairs at the $\psi(3770)$ resonance, a search has been performed for flavor-changing neutral current (FCNC) and lepton-number-violating (LNV) decays of $D^+$ mesons to final states with dielectrons [16]. No signal is observed and the 90% confidence level upper limits are $\mathcal{B}(D^+ \to \pi^+e^+e^-) < 7.4 \times 10^{-6}$, $\mathcal{B}(D^+ \to \pi^-e^+e^+) < 3.6 \times 10^{-6}$, $\mathcal{B}(D^+ \to K^+e^+e^-) < 6.2 \times 10^{-6}$, and $\mathcal{B}(D^+ \to K^-e^+e^+) < 4.5 \times 10^{-6}$.

---

[*]Speaker.
[†]I would like to thank my CLEO colleagues and the CESR staff

Searches for rare-decay processes have played an important role in the development of the Standard Model (SM). The absence of flavor-changing neutral currents (FCNC) in kaon decays led to the prediction of the charm quark [1], and the observation of $B^0$-$\bar{B}^0$ mixing, a FCNC process, signaled the very large top-quark mass [2]. This paper presents the results of searches for the FCNC decays [3, 4] $D^+ \to \pi^+ e^+ e^-$ and $D^+ \to K^+ e^+ e^-$, and the lepton-number-violating (LNV) decays [5] $D^+ \to \pi^- e^+ e^+$ and $D^+ \to K^- e^+ e^+$. (Charge-conjugate modes are implicit throughout this paper.) Short-distance FCNC processes in charm decays are much more highly suppressed by the GIM mechanism [6] than the corresponding down-type quark decays because of the range of masses of the up-type quarks. Observation of $D^+$ FCNC decays could therefore provide indication of non-SM physics or of unexpectedly large rates for long-distance SM processes like $D^+ \to \pi^+ V$, $V \to e^+ e^-$, with a real or virtual vector meson $V$. The LNV decays $D^+ \to \pi^- e^+ e^+$ and $D^+ \to K^- e^+ e^+$ are forbidden in the SM. Past searches have set upper limits for the four dielectron decay modes in our study that are of order $10^{-4}$ [7]. The limits for corresponding dimuon modes are about an order of magnitude more stringent.

The CLEO-c detector [8, 9, 10, 11] was used to collect a sample of 1.8 million $e^+ e^- \to \psi(3770) \to D\bar{D}$ events (1.6 million $D^{\pm}$ mesons) from an integrated luminosity of 281 pb$^{-1}$ provided by the Cornell Electron Storage Ring (CESR). CLEO-c consists of a six-layer low-mass drift chamber, a 47-layer central drift chamber, a ring-imaging Cherenkov detector (RICH), and a cesium iodide electromagnetic calorimeter, all operating inside a 1.0-T magnetic field. Background processes and the efficiency of signal-event selection are estimated with a GEANT-based [12] Monte Carlo (MC) simulation program. Physics events are generated by EvtGen [13] and final-state radiation (FSR) is modeled by the PHOTOS [14] algorithm. Signal events are generated with a phase-space model as a first approximation of non-resonant FCNC and LNV decays.

Candidate signal decays are reconstructed from well-measured charged-particle tracks that are consistent in three dimensions with production at the $e^+ e^-$ collision point. Electrons with momenta of at least 200 MeV are identified with a likelihood ratio that combines $E/p$, $dE/dx$, and RICH information. Charged kaons and pions with momenta of 50 MeV or greater are selected based on $dE/dx$ and RICH information. For each candidate decay of the form $D^+ \to h^{\pm} e^{\mp} e^+$, where $h$ is either $\pi$ or $K$, the energy difference $\Delta E = E_{\text{cand}} - E_{\text{beam}}$ is computed and the beam-constrained mass difference $\Delta M_{\text{bc}} = \sqrt{E_{\text{beam}}^2 - |\vec{p}_{\text{cand}}|^2} - M_{D^+}$, where $E_{\text{cand}}$ and $\vec{p}_{\text{cand}}$ are the measured energy and momentum of the $h^{\pm} e^{\mp} e^+$ candidate, $E_{\text{beam}}$ is the beam energy, and $M_{D^+}$ is the nominal mass of the $D^+$ meson [7]. The resolution for these quantities is improved by recovering bremsstrahlung photons that are detected in the calorimeter within 100 mrad of electron trajectories. This provides a signal-efficiency increase of $13\% - 18\%$, depending on decay mode.

Events with $D^+$ candidates satisfying $-30$ MeV $\leq \Delta M_{\text{bc}} < 30$ MeV and $-100$ MeV $\leq \Delta E < 100$ MeV are selected for further study. Within this region we define the "signal box" to be $-5$ MeV $\leq \Delta M_{\text{bc}} < 5$ MeV and $-20$ MeV $\leq \Delta E < 20$ MeV, corresponding to $\pm 3\sigma$ in each variable, as determined by MC simulation. The remainder of the candidate sample was used to assess backgrounds.

The expected branching fraction for the long-distance decay $D^+ \to \pi^+ \phi \to \pi^+ e^+ e^-$ is within the sensitivity of this analysis ($\sim 10^{-6}$). Candidates for this decay are selected based on the mass squared of the final-state $e^+ e^-$ (equal to the $q^2$ of the decay), with 0.9973 GeV$^2 \leq m_{e^+ e^-}^2 <$ 1.0813 GeV$^2$ defining the $\phi$-resonant region. This region is used both to veto the long-distance $D^+ \to \phi\pi^+ \to \pi^+ e^+ e^-$ contribution and to measure its branching fraction.

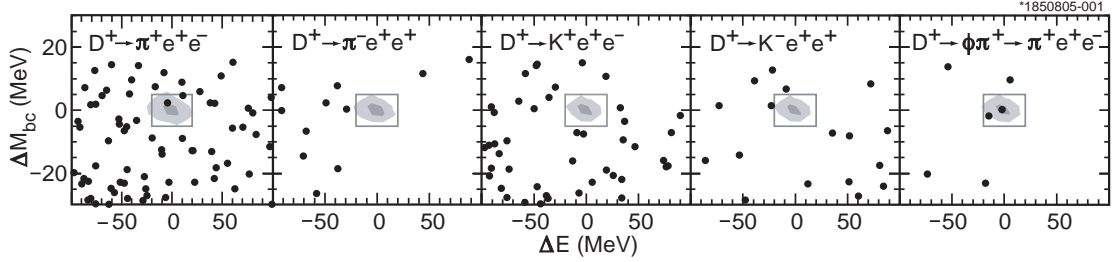

**Figure 1:** Scatter plots of $\Delta M_{bc}$ vs. $\Delta E$ obtained from data for each decay mode. The signal region, defined by $-20$ MeV $\leq \Delta E < 20$ MeV and $-5$ MeV $\leq \Delta M_{bc} < 5$ MeV, is shown as a box. The two contours for each mode enclose regions determined with signal MC to contain $50\%$ and $85\%$ of signal events, respectively.

Backgrounds in the $D^+ \to h^\pm e^\mp e^+$ candidate sample arise from both $D\bar{D}$ and non-$D\bar{D}$ sources. In $D\bar{D}$ events double semileptonic decays are dominant. Non-peaking backgrounds also arise from $\gamma$-conversion and Dalitz decays of $\pi^0$ and $\eta$.

Signal-selection and background-suppression criteria were optimized using MC simulation by minimizing the sensitivity variable

$$\mathcal{S} = \frac{\sum_{n=0}^{\infty} \mathcal{C}(n;N) \cdot \mathcal{P}(n;N)}{\epsilon \cdot N_{D^+}(\mathcal{L})}, \tag{1}$$

where $n$ is the observed number of events, $N$ is the expected number of background events, $\mathcal{C}$ is the 90% confidence coefficient upper limit on the signal, $\mathcal{P}$ is the Poisson probability, $N_{D^+}$ is the number of charged $D$ mesons (as a function of integrated luminosity $\mathcal{L}$), and $\epsilon$ is the signal-selection efficiency. $\mathcal{S}$ represents the average upper limit on the branching fraction that would be obtained from an ensemble of experiments if the true mean for the signal were zero. Sideband studies demonstrate that the MC simulation provides a good description of background events.

The decay mode $D^+ \to \pi^+ e^+ e^-$ is susceptible to background from $D^+ \to K_S^0 e^+ \nu_e$ accompanied by a semileptonic decay of the other $D$. This is suppressed by rejecting candidates when the signal $\pi^+$ and an oppositely charged track combine to give a mass $M_{\pi^+\pi^-}$ that satisfies $-5$ MeV $\leq M_{\pi^+\pi^-} - M_{K_S^0} < 5$ MeV, where $M_{K_S^0}$ is the nominal $K_S^0$ mass [7].

The residual background and the efficiencies after application of all selection criteria have been determined by MC simulation and are given for the four signal modes in Table 1. The model used to describe FCNC and LNV decays is phase space. The efficiency is observed to be quite uniform over the Dalitz plot, with the exception of the two corners at low $m_{ee}^2$, which are depleted by the 200-MeV minimum-momentum requirement for electron identification.

Scatter plots of $\Delta M_{bc}$ vs. $\Delta E$ for data events surviving all other cuts are shown in Fig. 1. For $D^+ \to \pi^+ e^+ e^-$, two events lie in the signal box, with an expected background of 1.99. For all other FCNC or LNV modes there are zero events in the signal box. With no evidence of a signal, 90% confidence level (CL) upper limits (UL) are calculated on the branching fraction for each mode from the observed number of events ($n$) in the signal box, the signal-detection efficiency ($\epsilon$), and the MC-estimated number of background events ($N$). The Poisson procedure [7] is used to calculate the 90% confidence level coefficient ($\mathcal{C}(n;N)$) upper limit on signal in the presence of expected background:

$$\text{UL} = \frac{\mathcal{C}(n;N)}{\epsilon \cdot (2 \cdot \sigma_{D^+D^-} \cdot \mathcal{L})}, \qquad (2)$$

where $\sigma_{D^+D^-}$ [15] is the $e^+ e^- \to \psi(3770) \to D^+ D^-$ cross section, $\mathcal{L}$ is the integrated luminosity, and $2 \cdot \sigma_{D^+D^-} \cdot \mathcal{L} = 1.6$ million is the number of charged $D$ mesons in our data sample. Results are given in Table 1: no evidence is found for either FCNC or LNV decays. The branching fraction for the resonant decay $D^+ \to \pi^+ \phi \to \pi^+ e^+ e^-$, is measured after finding two events in the signal region with an expected background of 0.04.

**Table 1:** Efficiencies ($\epsilon$), background estimates ($N$), observed yields ($n$), combined systematic uncertainties ($\sigma_{\text{syst}}$), and branching fraction results for four FCNC and LNV decay modes and for the resonant decay $D^+ \to \pi^+ \phi \to \pi^+ e^+ e^-$. Branching-fraction UL values are all at 90% CL.

| Mode | $\epsilon$ (%) | $N$ | $n$ | $\sigma_{\text{syst}}$ (%) | $\mathcal{B}$ ($10^{-6}$) |
|---|---|---|---|---|---|
| $\pi^+ e^+ e^-$ | 36.41 | 1.99 | 2 | 8.7 | < 7.4 |
| $\pi^- e^+ e^+$ | 43.85 | 0.48 | 0 | 7.1 | < 3.6 |
| $K^+ e^+ e^-$ | 26.18 | 1.47 | 0 | 10.0 | < 6.2 |
| $K^- e^+ e^+$ | 35.44 | 0.50 | 0 | 7.2 | < 4.5 |
| $\pi^+ \phi(e^+ e^-)$ | 46.22 | 0.04 | 2 | 7.4 | $2.8 \pm 1.9 \pm 0.2$ |

Systematic uncertainties in these results can be divided into two categories, background estimations and signal efficiencies.

Sources of uncertainties that are common to all results are the number of $D^+$ ($-3.2\%, +4.5\%$), tracking ($\pm 1\%$ per track or $\pm 3\%$ total), PID ($\pm 2.3\%$), and FSR ($\pm 4.0\%$ for $\pi^\pm e^\mp e^+$, $\pm 3.3\%$ for $K^+ e^+ e^-$, $\pm 3.5\%$ for $K^- e^+ e^+$, and $\pm 4.4\%$ for $\pi^+ \phi \to \pi^+ e^+ e^-$, estimated by comparing the efficiency before and after bremsstrahlung recovery).

Uncertainties in signal efficiency due to background-suppression cuts are estimated by comparing the efficiency before and after the cuts are applied: $\pm 5.2\%$ ($\pi^+ e^+ e^-$), $\pm 1.1\%$ ($\pi^- e^+ e^+$), $\pm 7.3\%$ ($K^+ e^+ e^-$), $\pm 1.0\%$ ($K^- e^+ e^+$), and $\pm 0.9\%$ ($\pi^+ \phi \to \pi^+ e^+ e^-$).

Uncertainty from using the phase-space model (as a first approximation for non-resonant decays) for the FCNC and LNV signal efficiency estimation is assessed by (somewhat arbitrarily) taking one quarter of the fraction of phase space which has non-uniform efficiency due to the electron identification momentum cut-off (200 MeV): $\pm 2.8\%$ ($\pi^\pm e^\mp e^-$) and $\pm 3.8\%$ ($K^\pm e^\mp e^-$).

For the results in Table 1, we increase the upper limits to account for systematic uncertainties by decreasing the efficiency by $1\sigma_{\text{syst}}$ (combined systematic uncertainty).

In summary, limits are set on four rare (FCNC) and forbidden (LNV) decays of charged $D$ mesons to three-body final states with dielectrons. No evidence of signals is found and 90% confi-dence level upper limits are:

$$\begin{aligned} \mathcal{B}(D^+ \to \pi^+ e^+ e^-) &< 7.4 \times 10^{-6} \\ \mathcal{B}(D^+ \to \pi^- e^+ e^+) &< 3.6 \times 10^{-6} \\ \mathcal{B}(D^+ \to K^+ e^+ e^-) &< 6.2 \times 10^{-6} \\ \mathcal{B}(D^+ \to K^- e^+ e^+) &< 4.5 \times 10^{-6} \end{aligned}$$

The results for these dielectron modes are significantly more restrictive than previous limits, and reflect sensitivity comparable to the searches for dimuon modes [7]. Due to the dominance of long-distance effects in FCNC modes, we separately measure the branching fraction of the resonant decay $D^+ \to \pi^+ \phi \to \pi^+ e^+ e^-$, obtaining $\mathcal{B}(D^+ \to \phi \pi^+ \to \pi^+ e^+ e^-) = (2.8 \pm 1.9 \pm 0.2) \times 10^{-6}$. This is consistent with the product of known world average [7] branching fractions, $\mathcal{B}(D^+ \to \phi \pi^+ \to \pi^+ e^+ e^-) = \mathcal{B}(D^+ \to \phi \pi^+) \times \mathcal{B}(\phi \to e^+ e^-) = [(6.2 \pm 0.6) \times 10^{-3}] \times [(2.98 \pm 0.04) \times 10^{-4}] = (1.9 \pm 0.2) \times 10^{-6}$.

We gratefully acknowledge the effort of the CESR staff in providing us with excellent luminosity and running conditions. This work was supported by the National Science Foundation and the U.S. Department of Energy.